\documentclass[prl,floatfix,twocolumn]{revtex4}

\usepackage{amsmath}
\usepackage{graphicx}

\begin{document}
\title{Comment on ``Saturation of the all-optical Kerr effect''}

\maketitle

In a recent Letter \cite{bree}, Br\'ee, Demircan, and Steinmeyer calculated higher-order Kerr coefficients $n_{2k}$ in noble gases using a generalized Kramers-Kronig (KK) relation applied to high order nonlinear processes. They found that the nonlinear index $\Delta n$ for argon obtained from $\Delta n = \sum n_{2k} I^k$ saturates and goes negative in the intensity range $I\sim$40-50 TW/cm$^2$, well below the threshold for ionization.

While a recent experiment has shown conclusively that there is no nonlinear index saturation below the ionization threshold \cite{wahlstrand_optical_2011}, here we show on theoretical grounds that the calculation leading Br\'ee \emph{et al.}~to their conclusion is incorrect. It errs by using multiphoton ionization (MPI) rates in a regime where perturbation theory breaks down.

The KK relation used by \cite{bree} is
\begin{equation}
n_{2k}(\omega)=\frac{\hbar\omega}{\pi} \mathcal{P} \int_0^\infty (\Omega + \omega) \frac{\sigma_{k+1} (\xi)}{\Omega^2-\omega^2} d\Omega,
\end{equation}
where $\sigma_k$ is the coefficient for $k^{\mathrm{th}}$ order MPI with rate $w_k=\sigma_k I^k$,  $\xi=(\Omega+k\omega)/(k+1)$, and $\mathcal{P}$ denotes principal value.
The coefficients $\sigma_k$ used by Br\'ee \emph{et al.}~(Eq.~(1) of \cite{bree}) are obtained from the $\gamma\rightarrow \infty$  or MPI limit of a recently derived ionization rate $R$ \cite{mod}, where, in atomic units, $\gamma=(2I_p)^{1/2}\omega/E_0$ is the Keldysh parameter and where $I_p$ is the ionization potential, $\omega$ is the laser frequency and $E_0$ is the peak laser field.
The rate $R$ (Eq. (6) of \cite{mod}) handles both the $\gamma\gg 1$ MPI limit and the $\gamma\ll 1$ tunneling limit, and provides a reasonably accurate interpolation for $\gamma$ lying in between.

In Fig.~1 we plot $R$ for argon ($I_p=15.76$ eV) for $\hbar\omega =1.55$ eV ($\lambda=800$ nm) as a function of intensity (red dashes). At low intensities $R\propto I^M$, where $M$ (=11 for Ar) is the minimum number of photons for ionization, and at higher intensities $R$ shows the characteristic roll off as the tunneling regime is approached.  We also plot the total MPI rate $w=\sum_{k=11}^{50} w_k$ (blue dots), where the contributions for $k>11$ correspond to above threshold ionization. Above $\sim70$ TW/cm$^2$ , $w$ diverges badly. However, even as low as 5 TW/cm$^2$, $R$ and $w$ differ by a factor of $\sim2$. Thus we immediately see that the terms containing $\sigma_k$ contributing to $w$ are quite inadequate to describe ionization in the $>40$ TW/cm$^2$ region where the novel behavior of $\Delta n$ is claimed.
Thus, these $\sigma_k$ are problematic for use in Eq.~(1).
Forcing a mathematical fit of $w$ to $R$, using the $\sigma_k$ as free parameters, would require some of the $\sigma_k$ to be negative for large $k$, an unphysical result.

\begin{figure}
\center{\includegraphics[width=8.5cm]{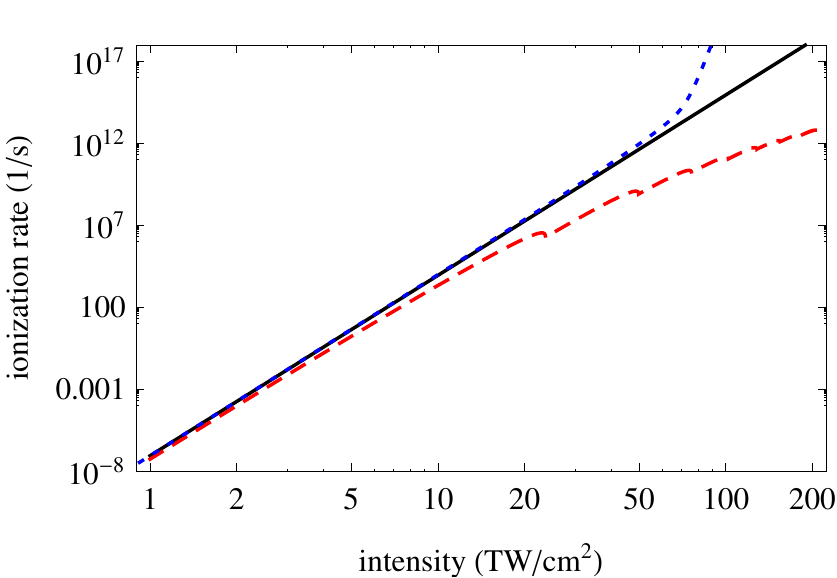}}
\caption{(color online) Ionization rates $w$ (blue dots), $w_k$ (solid black), and $R$ (red dashes) calculated for Ar using the models in \cite{bree} and \cite{mod}. Here $w_k$ is the $k^{\mathrm{th}}$ order MPI rate applicable for $\gamma\gg 1$ (shown for $k=11$), $w$ is the sum over $w_k$ from $k=11$ to 50, and $R$ is the rate from \cite{mod} applicable from the MPI though tunneling regimes ($\gamma\ll 1$).}
\label{ionfig}
\end{figure}

We now show how Br\'ee \emph{et al.}~obtain saturation and negative excursion of $\Delta n$.
In \cite{bree} it is stated that $n_{2k}>0$ for $k\leq9$ and $n_{2k}<0$ for $k\geq10$.
It is the $k\geq10$ contributions which cause $\Delta n$ to saturate and go negative.
First, we note that it is the $0<\Omega<\omega$ portion of the KK integral (1) which can contribute negatively to $n_{2k}$. In that interval, $\xi$ in $\sigma_{k+1}(\xi)$ ranges from $\omega k/(k+1)$ to $\omega$. However, $\sigma_{k+1} (\xi) = 0$ for $\xi<\xi_{\mathrm{thresh}}=I_p/[(k+1)\hbar]$, as appropriate for a MPI coefficient \cite{mod}.
Suppose $k=1$. Then the KK integral would contribute negatively for $\omega/2<\xi<\omega$, except for the fact that $\sigma_2(\xi)=0$ in this range.
Thus $n_2>0$. For the same reasons, $n_{2k}>0$ for $k\leq 9$.
However, for $k=10$, the KK integral contributes negatively in the interval $(10/11)\omega<\xi<\omega$ because $\xi\sim\xi_{\mathrm{thresh}}$ and $\sigma_{11}(\xi)>0$.
Similar negative contributions occur for $k>10$, where $\xi>\xi_{\mathrm{thresh}}$.
Thus, \emph{all} negative Kerr coefficients $n_{2k}$ for $k\geq10$ are computed using $\sigma_{k+1}(\xi)$ for $\xi\sim\omega$, a region for which $\gamma\sim1.8$ at 40 TW/cm$^2$ in Ar.
In using Eq.~(1) to calculate the terms $n_{2k}$ in $\Delta n=\sum n_{2j} I^j$, $\gamma$ has therefore been implicitly sampled from $\gamma=10$ ($k=1$) to $\gamma=1.8$ ($k=10$), a range over which an underlying theory of ionization should encompass the transition from MPI to tunneling.
Instead, Br\'ee \emph{et al.}~have used MPI exclusively, as appropriate for a perturbative expansion. It is not a coincidence that $n_{2k}<0$ for $k\geq10$: this is the $\lambda=800$ nm MPI threshold for Ar, where $\sigma_{11}(\omega)>0$. 
Essentially, the saturated and negative intensity dependence of $\Delta n$ obtained by Br\'ee \emph{et al.}~is an artifact of a perturbation expansion in a laser intensity region where it is far from appropriate.
We note that Br\'ee \emph{et al.}~justify their use of rates in the MPI limit by quoting $\gamma=1.62$ at 50 TW/cm$^2$.
While this might be an adequate assumption if the rates are to be used only for order-of-magnitude estimates of ionization yields, it is clearly incorrect for calculating nonlinear coefficients using KK theory.

Finally, we note the remarkable coincidence that an experiment \cite{loriot} and a theory \cite{bree} both show apparent saturation of the Kerr effect in the same intensity range for completely different reasons. In the experiment the saturating and negative phase shift was likely caused by diffraction from a plasma grating \cite{wahlstrand_effect_2011}; here it is caused by the extension of a perturbative theory of ionization to an intensity range where it breaks down.

\bigskip

J. K. Wahlstrand and H. M. Milchberg

\small Institute for Research in Electronics and Applied Physics, University of Maryland, College Park, MD 20742

\end{document}